\documentclass{PoS}

\usepackage{latexsym}
\usepackage{amssymb}
\usepackage{amsbsy}
\usepackage{epsfig}

\newcommand{\om}{\omega}
\newcommand{\be}{\begin{equation}}
\newcommand{\ee}{\end{equation}}
\newcommand{\bea}{\begin{eqnarray}}
\newcommand{\eea}{\end{eqnarray}}
\newcommand{\nn}{\nonumber}

\newcommand{\bra}{\langle}
\newcommand{\ket}{\rangle}
\newcommand{\xv}{{\mathbf x}}
\newcommand{\jv}{{\mathbf j}}

\newcommand{\id}{1\!\!\!1}

\title{Transport and spectral functions in high-temperature 
QCD\thanks{Plenary talk}}

\ShortTitle{Transport and spectral functions in high-temperature QCD}

\author{
Gert Aarts\\
Department of Physics, Swansea University\\
Swansea, SA2 8PP, United Kingdom\\
        E-mail: \email{g.aarts@swan.ac.uk}}

\abstract{

The current status of transport coefficients in relativistic field 
theories at high temperature is reviewed. I contrast weak coupling results 
obtained using kinetic theory/diagrammatic techniques with strong coupling 
results obtained using gauge/gravity duality, and describe the recent 
developments in extracting transport coefficients and spectral functions 
from lattice QCD simulations. The fate of quarkonium at high temperature 
as seen from the lattice is briefly mentioned as well.

}

\FullConference{The XXV International Symposium on Lattice Field Theory\\
		 July 30-4 August 2007\\
		 Regensburg, Germany}

\begin{document}

\section{Introduction}

Transport coefficients characterize fluctuations and relaxation on long 
length and time scales in systems slightly away from thermal equilibrium.
 Consider a conserved current, $\partial_t n + \nabla\cdot \jv = 0$, in a 
system with a small nonuniformity in the density $n(t,\xv)$. To first 
approximation, a current will form to wash out the imbalance: $\jv =-D 
\nabla n +\ldots$, where the dots indicate higher order terms in 
gradients. Combining the conservation law with this constitutive equation 
immediately yields the diffusion equation $\partial_t n = D\nabla^2 n$. 
The proportionality factor $D$ is the diffusion constant and is the only 
quantity determined by the underlying microscopic theory: one may view it 
as a {\em low-energy constant}.
 Similarly, consider the conserved energy-momentum tensor $T^{\mu\nu}$. 
While in equilibrium $T^{ij}=\delta^{ij}P$, with $P$ the 
pressure, a perturbation characterized by a nonuniform flow field 
$u(t,\xv)$ will change this to (in the local rest frame, where $T^{0i}=0$)
 \be
 T^{ij}=\delta^{ij}P - \eta\left(
\partial^iu^j +
\partial^ju^i -
\frac{2}{3}\delta^{ij}\partial_lu^l \right) - 
\zeta \delta^{ij} \partial_lu^l +\ldots.
\ee
 Again the dots indicate higher order terms in gradients. The coefficient 
of the traceless combination is the shear viscosity  $\eta$ and $\zeta$ is 
the bulk viscosity. Combining energy-momentum conservation with the 
constitutive equation above yields the hydrodynamic equations, predicting 
e.g.\ sound waves.
 In the context of thermal QCD one may therefore view hydrodynamics as the 
{\em low-energy effective theory} describing real-time dynamics at 
sufficiently large length and time scales. The form of the hydrodynamic 
equations is fixed by combining exact conservation laws and constitutive 
equations, which are obtained in a gradient expansion. In the latter, a 
number of {\em low-energy constants} appear, determined by QCD: shear 
viscosity $\eta$, bulk viscosity $\zeta$, electrical conductivity 
$\sigma$, diffusion constants $D$, etc. In this talk I review the progress 
in determining these coefficients from first principles.

The recent interest in transport in QCD and related theories is mainly due 
to the relativistic heavy ion program at RHIC \cite{Arsene:2004fa}. The 
remarkable effectiveness of ideal hydrodynamics in describing heavy ion 
phenomenology \cite{Kolb:2000fha,Teaney:2000cw} suggests that transport 
coefficients are very small (when appropriately normalized). This in turns 
implies that thermalization times are short and interactions are strong, 
suggesting that in the temperature range $1\lesssim T/T_c\lesssim 2$ the 
quark-gluon plasma is not a weakly coupled system of quarks and gluons, 
but instead strongly interacting (sQGP).
 A second reason for interest in transport is the fertile applicability of 
gauge/gravity duality (or AdS/CFT correspondence) to study strongly 
coupled thermal gauge theories in the hydrodynamic regime 
\cite{Son:2007vk}. This has led to many (semi-)analytical results for 
those theories that admit a gravity dual and provides an important 
stimulus for QCD, where a gravity dual is not available. The best-known 
example concerns the ratio of the shear viscosity and the entropy density 
$s$,
 \be
 \frac{\eta}{s} = \frac{1}{4\pi}, 
\ee
 which is obtained in all thermal gauge theories in the (strongly coupled) 
regime described by a dual gravity theory. This ratio is much smaller than 
in weakly coupled QCD, where 
 \be
 \label{eqetasweak}
 \lim_{g\to 0}\, \frac{\eta}{s} \sim \frac{1}{g^4\ln 1/g} \to \infty.
\ee
 It is an open question what this ratio is in QCD just above the 
deconfinement transition. Below I describe recent progress in lattice QCD 
that will bring us closer to answering that question.

\section{Transport coefficients at weak and strong coupling}

According to the Kubo relations, transport coefficients are given by the 
slope of current-current spectral functions, computed in thermal 
equilibrium, at vanishing energy. For example, the electrical conductivity 
$\sigma$ (or charge diffusion constant $D$ times susceptibility 
$\Xi$) is determined by
 \be
 \sigma = D\, \Xi = \lim_{\om\to 0} \frac{\rho^{11}(\om)}{2\om},
\ee
 where the current-current spectral function is given by
 \be 
\label{eqrho2}
\rho^{\mu\nu}(\om) = \int d^4x\, e^{i\om t} \bra [ 
j^\mu(t,\xv),j^\nu(0)]\ket_{\rm eq}.
 \ee
For a single fermionic charge carrier with charge $e$ the 
electromagnetic current reads $j^\mu = e\bar\psi\gamma^\mu\psi$. 
Similarly, the shear and bulk viscosities are given by
\be
 \eta = \lim_{\om\to 0} \frac{\rho^{12,12}(\om)}{2\om},
 \;\;\;\;\;\;\;\;
 \;\;\;\;\;\;\;\;
 \;\;\;\;\;\;\;\;
 \zeta = \frac{1}{9}\lim_{\om\to 0} 
\frac{\rho^{ii,jj}(\om)}{2\om},
\ee
in terms of the energy-momentum tensor spectral function
\be
\label{eqrho4}
\rho^{\mu\nu,\rho\sigma}(\om) = \int d^4x\, e^{i\om t} \bra [
T^{\mu\nu}(t,\xv),T^{\rho\sigma}(0)]\ket_{\rm eq}.
\ee

\subsection{Weak coupling} 

At weak coupling transport coefficients can be computed using either kinetic 
theory or by summing sets of Feynman diagrams. In this limit, 
transport coefficients are manifestly proportional to the mean free path 
or the inverse collisional width $1/\Gamma$. They are therefore very large 
and inversely proportional to the coupling constants in the theory, see 
Eq.\ (\ref{eqetasweak}). In ultrarelativistic QCD the shear viscosity, 
electrical conductivity and diffusion constants have been computed to 
leading-logarithm order \cite{Arnold:2000dr} and subsequently to full 
leading order \cite{Arnold:2003zc} in the gauge coupling.
 The relevant physics at leading-log are those $2\leftrightarrow 2$ 
scattering processes that are logarithmically sensitive to infrared 
screening effects.
 The extension to full leading order requires, besides the other 
$2\leftrightarrow 2$ scattering processes, also the inclusion of specific 
particle number changing processes \cite{Arnold:2002zm}.
 The bulk viscosity is more complicated since particle number changing 
processes need to be included from the start; the first calculation in QCD 
can be found in Ref.\ \cite{Arnold:2006fz}. It is shown that a 
parametrically correct estimate is given by
 \be
 \label{eqbulk}
 \zeta \approx 15\eta \left(1/3 - v_s^2\right)^2,
 \ee
 where for light quarks the speed of sound $v_s$ is determined by $v_s^2 - 
1/3 \sim  
\beta(g^2) \sim g^4$.
 The equivalence between kinetic theory and diagram summation in thermal 
gauge theories has been demonstrated in a number of papers 
\cite{Valle Basagoiti:2002ir,Aarts:2002tn,Aarts:2003bk,Gagnon:2006hi,Carrington:2007ea}.
A one-loop calculation is never sufficient; instead an infinite set of 
ladder diagrams has to be taken into account.

 A complete leading-order calculation is also possible in the large $N_f$ 
limit, where $N_f$ indicates the number of flavours. This has been done 
using kinetic theory for massless quarks \cite{Moore:2001fg} and extended 
to massive quarks using diagrams \cite{Aarts:2005vc}. Only Coulomb 
scattering processes contribute. Also in this limit one is effectively in 
the weakly coupled regime, since $\eta/s\sim N_f\to \infty$.

\begin{figure}
\centerline{\psfig{figure=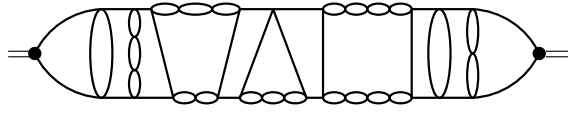,width=7.5cm}}
 \caption{Typical skeleton diagram contributing to the shear 
viscosity in the $O(N)$ model at large $N$.
 }
\label{figladder}
\end{figure}

\begin{figure}[b]
\centerline{\psfig{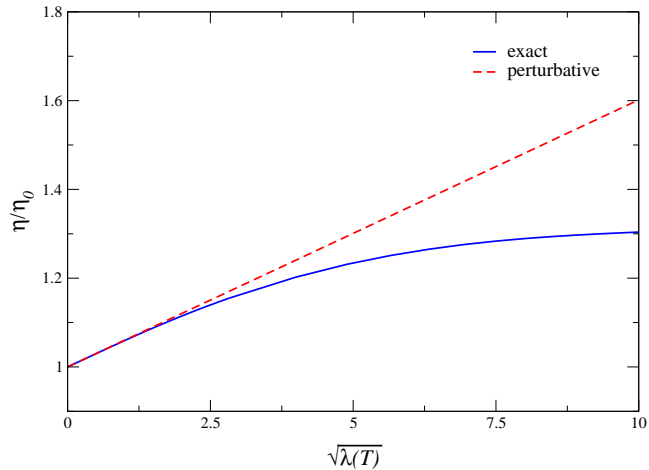}}
 \caption{Shear viscosity $\eta$, normalized with the result at vanishing 
coupling, in the $O(N)$ model as a function of $\sqrt{\lambda}$: 
comparison between the perturbative \cite{Moore:2007ib} and the exact
\cite{Aarts:2004sd} result in the large $N$ limit.
 }
\label{figguy}
\end{figure}

In all cases above only the leading-order result is known. This prevents 
any reasonable extrapolation to stronger coupling, unlike in the case of 
e.g.\ the pressure where the terms up to $g^6\ln 1/g$ are known 
\cite{Kajantie:2002wa}. Recently the first subleading correction to a 
transport coefficient in a relativistic field theory has been computed by 
Moore \cite{Moore:2007ib}, namely to the shear viscosity in scalar 
$\phi^4$ theory.
 The shear and bulk viscosity for single-component scalar theories have
been computed a long time ago to leading order at weak coupling using
ladder diagram summation \cite{Jeon:1994if}. In the $O(N)$ model the shear
viscosity has also been found to leading order in the $1/N$ expansion
\cite{Aarts:2004sd}, using similar techniques as in large $N_f$ gauge
theories. It is shown in Ref.\ \cite{Moore:2007ib} that the first
subleading correction to the shear viscosity at weak coupling is sensitive
to soft physics and can be extracted from the same set of diagrams that
contribute at leading order by an expansion in terms of the thermal mass
$m_{\rm th}/T \sim \sqrt{\lambda}$. The result in the $O(N)$ model, with a
$\lambda/(4!N) (\phi_a\phi_a)^2$ interaction, reads \cite{Moore:2007ib}
 \bea
 \nn
 \eta &=&  \frac{T^3}{\lambda^2}\frac{3N^3}{N+2}
 \left( 3033.54 + 1548.3 \sqrt{\left(1+\frac{2}{N}\right) 
\frac{\lambda}{72}} 
  + {\cal O}(\lambda) \right)
\\
  &=& \frac{3N^2T^3}{\lambda^2} 
 \left( 3033.54 + 1548.3 \sqrt{ \frac{\lambda}{72}}
  + {\cal O}(\lambda, 1/N) \right), 
 \eea 
 where the second line is valid in the combined large $N$ and weak 
coupling limit. As mentioned above, the complete large $N$ result, without 
employing the weak-coupling expansion, is also known \cite{Aarts:2004sd}. 
This includes a resummation of the thermal mass to all orders, as well as 
the inclusion of other diagrams suppressed in the weak coupling limit, see 
Fig.\ \ref{figladder}. 
Therefore, an assessment of the perturbative expansion can now 
be made. This is shown in Fig.\ \ref{figguy}, where the perturbative 
result $\eta/\eta_0 = 1 + \# \sqrt{\lambda}$ is compared with the complete 
large $N$ result. Here $\eta_0$ is the result at vanishing coupling, which 
is used for normalization. From the first two terms, it appears that the 
convergence of the weak coupling expansion is not impressive.
 A possible approach to improve the convergence \cite{Moore:2007ib} would 
be to include the thermal mass without re-expanding, as has been tried for 
the pressure \cite{Karsch:1997gj}. How to include all ${\cal O}(\lambda)$ 
effects is currently not known.

 Considerable effort has been spent on nonrelativistic dynamics and
diffusion of heavy quarks (with $M_q\gg T$). I will not discuss that here,
but refer to Ref.\ \cite{Moore:2004tg} where definitions and a
leading-order calculation of the momentum diffusion coefficient can be
found, and to Ref.\ \cite{CaronHuot:2007gq} for the first next-to-leading
order result at weak coupling.

\subsection{Strong coupling}

In order to complement weak coupling results as discussed above, it is
desirable to have an analytic method tailored for strongly coupled field
theories. For a certain class of thermal gauge theories such a method is
available, via the gauge/gravity or AdS/CFT correspondence
\cite{Maldacena:1997re}. It uses the duality between a field theory at
finite temperature and black holes in AdS space described in string
theory. The best-known case relates ${\cal N}=4$ supersymmetric Yang-Mills
theory to type IIB string theory on AdS$_5\times S^5$, but there are many
other field theories for which a gravity dual can be found (but not QCD).
The parameters in field theory, the number of colours $N_c$ and the
coupling constant $g^2$, are related to the parameters of string theory in
AdS space. It turns out that the duality is most powerful when $g^2$ is
small, since then loops in the string theory are suppressed, but with the
't Hooft coupling $\lambda=g^2N_c$ large, since then stringy effects are
suppressed and the string theory reduces to a supergravity theory.
Strongly coupled gauge theories, with large 't Hooft coupling, are
therefore the natural area of application. Using the gauge/gravity
correspondence, it has been shown that the field theories for which the
duality holds behave hydrodynamically at large length and time scales,
which supports the framework of (nearly ideal) hydrodynamics to understand
the dynamics in thermal gauge theories at strong coupling.

The most famous example concerns the shear viscosity in ${\cal N}=4$ SYM. 
In the limit that both $N_c$ and $\lambda$ go to infinity, it is 
equal to 
\cite{Policastro:2001yc}
 \be
 \eta = \frac{\pi}{8}N_c^2 T^3.
 \ee
 In the same limit the entropy density is \cite{Gubser:1996de}
\[
 s = \frac{\pi^2}{2} N_c^2 T^3 = \frac{3}{4} s\Big|_{\lambda=0},
\]
 such that the ratio is $\eta/s=1/4\pi$. This result for the ratio is 
universal and is achieved in all thermal gauge theories that can be 
described by a gravity dual \cite{Kovtun:2004de,Buchel:2003tz}. The bulk 
viscosity vanishes due to conformal invariance.

The shear viscosity in ${\cal N}=4$ SYM at strong coupling is another rare
example of a transport coefficient where the first subleading correction
is known. One finds \cite{Buchel:2004di} 
 \be
 \eta = \frac{\pi}{8}N_c^2 T^3 \left( 1+ 
 \frac{75\zeta(3)}{4\lambda^{3/2}} + \ldots \right).
\ee
The first correction to the ratio then reads
\be
 \frac{\eta}{s} = \frac{1}{4\pi} \left( 
 1+\frac{135\zeta(3)}{8\lambda^{3/2}} + \dots \right).
\ee
 It has been conjectured that $1/4\pi$ is a lower bound for a wide 
class of systems, including e.g.\ the quark-gluon plasma and trapped 
atomic gases \cite{Kovtun:2003wp,Kovtun:2004de}. For a recent critical 
assessment, see Ref.~\cite{Cherman:2007fj}.

Other transport coefficients that have been computed include the 
$R$-charge diffusion coefficient and the bulk viscosity. These are not 
universal and depend on the theory under consideration. In ${\cal N}=4$ 
SYM the $R$-charge diffusion constant, susceptibility and conductivity are 
given by \cite{Policastro:2002se}
 \be
 D=\frac{1}{2\pi T},
 \;\;\;\;\;\;\;\;\;\;\;\;\;\;\;\;\;\;\;
 \Xi = \frac{N_c^2T^2}{8},
 \;\;\;\;\;\;\;\;\;\;\;\;\;\;\;\;\;\;\;
 \sigma = \frac{N_c^2T}{16\pi}.
 \ee
The bulk viscosity in a number of theories is discussed in Ref.\ 
\cite{Buchel:2007mf}. For strongly coupled systems it is proposed that it 
satisfies (cf.\ Eq.\ (\ref{eqbulk}) and the absent square)
 \be
 \zeta \geq 2\eta \left(1/3 - v_s^2\right).
 \ee
 Also heavy quark dynamics has been studied extensively in strongly 
coupled ${\cal N}=4$ SYM, see e.g.\ Ref. \cite{Herzog:2006gh} for a clear 
discussion.

 Due to the interest in ${\cal N}=4$ SYM as a testbed for thermal field
dynamics, various quantities have also been studied at weak coupling.  The
ratio $\eta/s$ has been computed in this theory to full leading order in
the small $\lambda$ limit (the result is independent of $N_c$) and
compared with the same quantity in QCD \cite{Huot:2006ys}.  Heavy quark
diffusion has been studied to leading order at weak coupling in Ref.\
\cite{Chesler:2006gr}.

\section{Spectral functions}

So far, in discussing transport coefficients, the focus has been on the 
zero-energy limit of spectral functions. In this section I give a 
(restricted) review of spectral functions at arbitrary $\om$, motivated 
mainly by hydrodynamic structure at weak and strong coupling. This will 
turn out to be useful for lattice QCD studies of spectral functions, to be 
discussed next.

\begin{figure}
\centerline{\psfig{figure=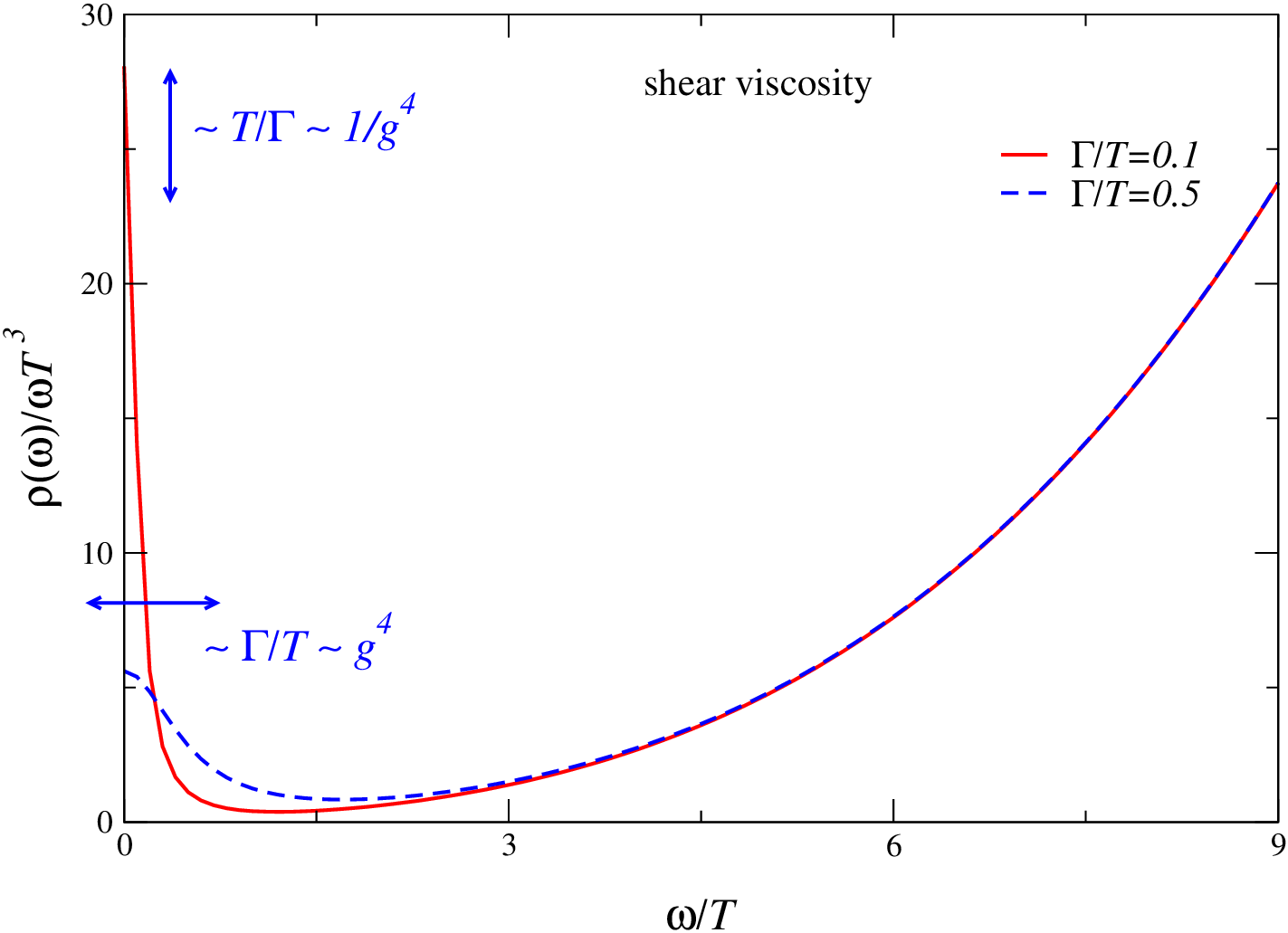,width=7.5cm}
\psfig{figure=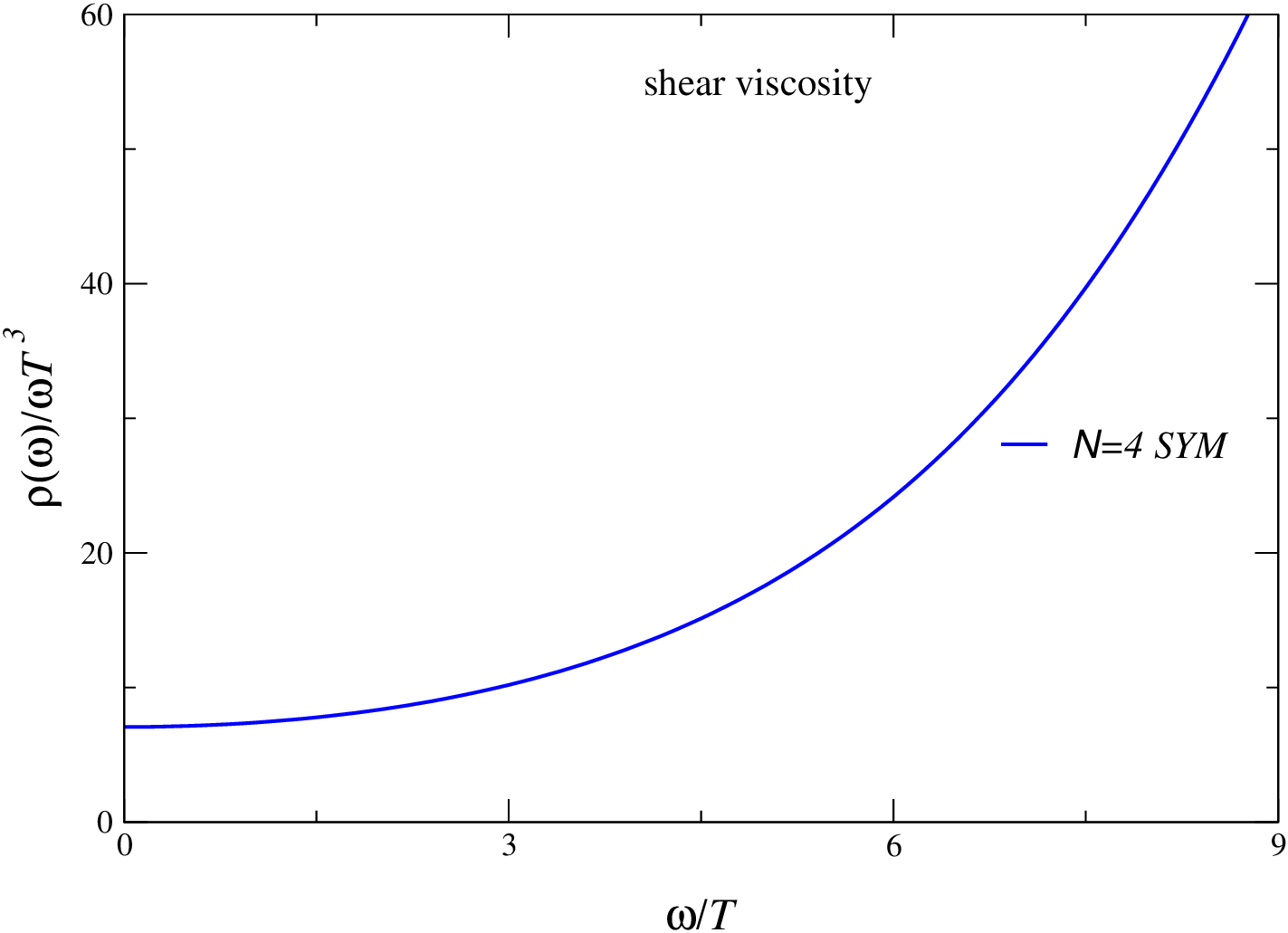,width=7.5cm}}
\caption{Shear viscosity: $\rho(\om)/\om T^3$ vs.\ $\om/T$ in 
QCD (sketched) for two values of the collisional width $\Gamma/T=0.1, 0.5$ 
(left) and in ${\cal N}=4$ SYM at strong coupling (right).
}
\label{figeta}
\vspace{0.5cm}
\centerline{\psfig{figure=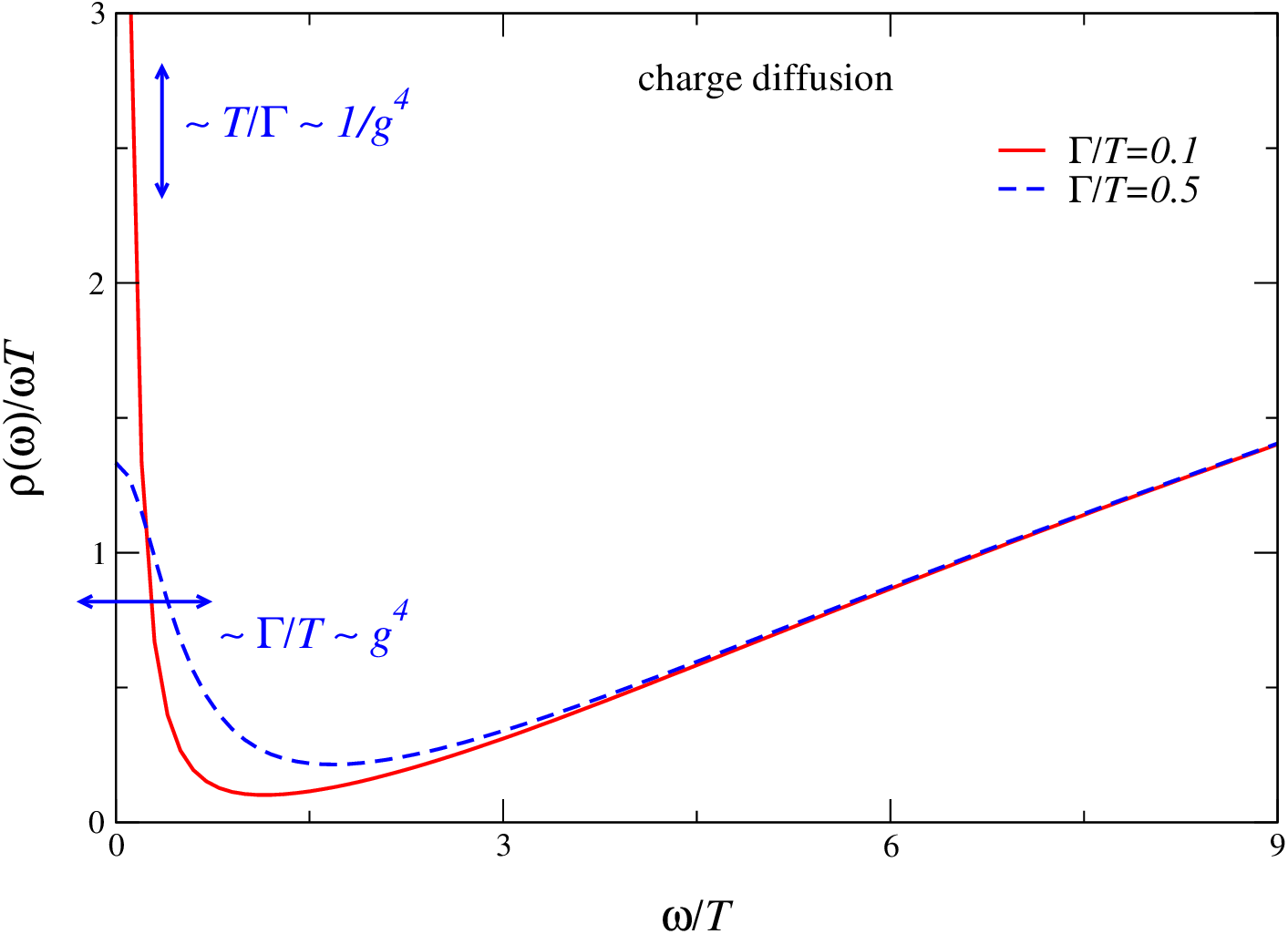,width=7.5cm}
\psfig{figure=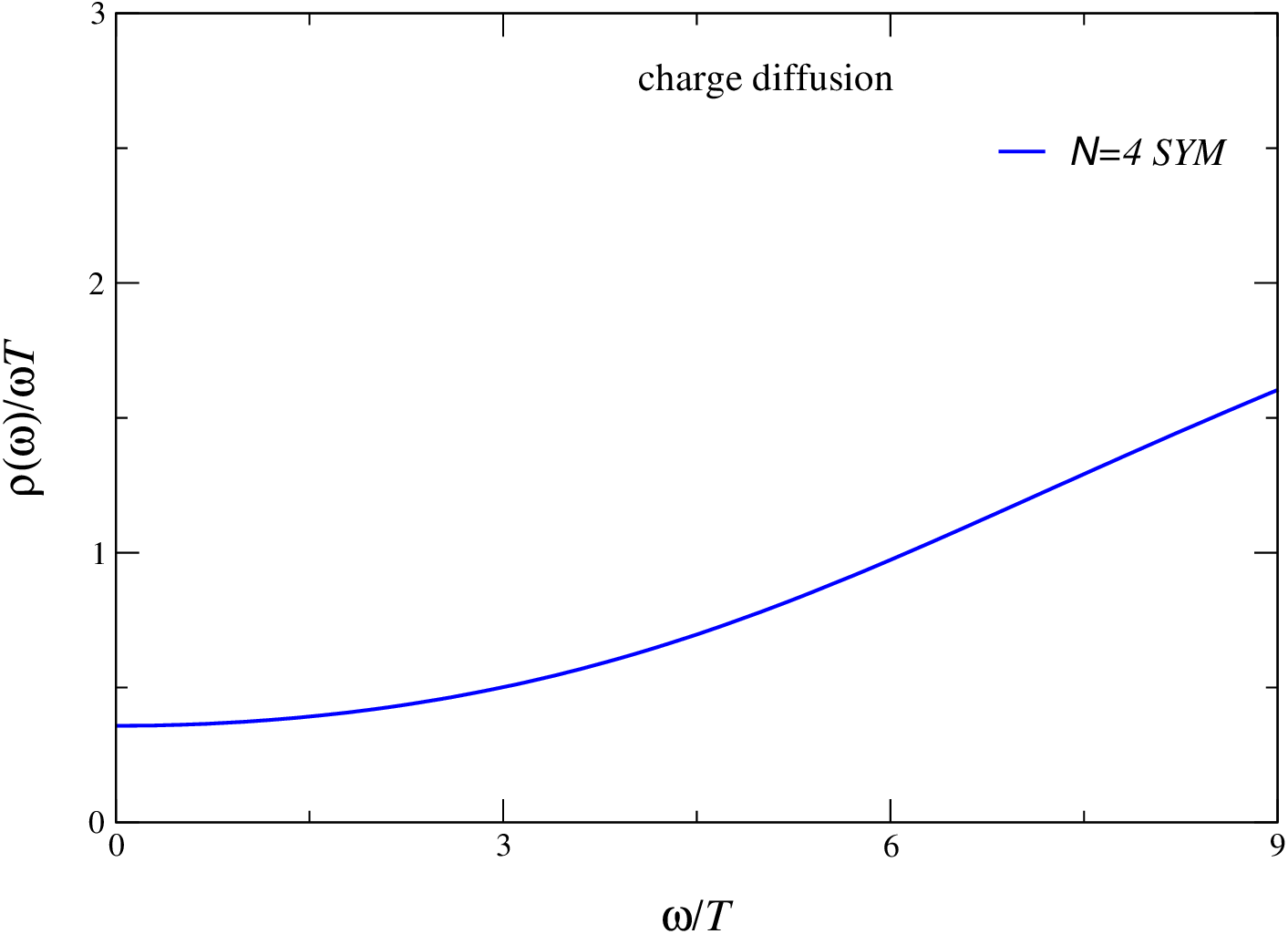,width=7.5cm}}
\caption{Charge diffusion: $\rho(\om)/\om T$ vs.\ $\om/T$ in QCD 
(sketched) for two values of the collisional width $\Gamma/T=0.1, 0.5$ 
(left) and in ${\cal N}=4$ SYM at strong coupling (right).
}
\label{figsig}
\end{figure}

In the weak coupling limit, current-current spectral functions have a 
characteristic energy dependence \cite{Aarts:2002cc}. At large energy they 
increase as $\om^n$, where the power $n$ is determined by the (mass) 
dimension. For example, the energy-momentum tensor spectral function 
(\ref{eqrho4}) increases as $\om^4$ and the EM current spectral function 
(\ref{eqrho2}) as $\om^2$ (unless there is a cancelation between 
components). At small energies, there is a transport peak. In free field 
theory, this peak manifests itself as a singular term,
 \be
 \frac{\rho(\om)}{\om} \sim 2\pi\delta(\om), 
 \ee
 reflecting that in a free theory the mean free path is infinite and 
transport coefficients diverge. Interactions regulate this singular 
behaviour and after the resummation of the collisional width $\Gamma$, the 
transport peak is modified to
 \be 
 \frac{\rho(\om)}{\om} \sim \frac{2\Gamma}{\om^2+\Gamma^2}.
 \ee
 For simplicity, I parametrized the effect of interactions with a single 
constant width $\Gamma$; in reality the width depends on the momentum of 
the (quasi-)particles contributing to transport. At weak coupling the 
collisional width $\Gamma\sim g^4T$: therefore the transport peak is 
narrow ($\sim g^4$) and high ($\sim 1/g^4$). 

The $\om$ dependence of spectral functions of conserved charges, such as 
${\cal E} = \int_\xv T^{00}(t,\xv)$ or $Q=\int_\xv j^0(t,\xv)$, is 
completely fixed by the conservation laws. For example, the spectral 
function for the total charge density reads
 \be
 \frac{\rho^{00}(\om)}{\om} = \Xi 2\pi \delta(\om),
 \ee
 where $\Xi$ is again the susceptibility. This ensures that unequal-time 
correlation functions, such $\bra Q(t)Q(0)\ket$, are in fact constant and 
equal to $VT\Xi$, where $V$ is the spatial volume and it is assumed that 
$\bra Q\ket=0$. Here I only discuss spectral functions at zero spatial 
momentum; for nonzero spatial momentum see e.g.\ Ref.\ 
\cite{Aarts:2005hg}.

 Combining the rising high-energy part and the transport peak at small
energies yields spectral functions as shown in Fig.\ \ref{figeta} (left)
for the energy-momentum tensor (in the channel relevant for the shear
viscosity) and Fig.\ \ref{figsig} (left) for the EM current (charge
diffusion). Here the width is taken as a free parameter and artificially
increased to larger values ($\Gamma/T\lesssim 1$), while keeping the
rest of the spectral function unchanged. As a result the transport peak,
which is narrow at weak coupling ($\Gamma/T\ll 1$), becomes less
singular and gets smeared out.

 Interestingly enough, this behaviour can be compared with spectral 
functions computed in ${\cal N}=4$ SYM in the limit of large $N_c$ and 
large 't Hooft coupling $\lambda$, using gauge/gravity duality 
\cite{Teaney:2006nc,Kovtun:2006pf}. The energy-momentum tensor spectral 
function has to be computed numerically by solving an ordinary 
differential equation, but for the $R$-current the analytical result is 
known and reads \cite{Myers:2007we}
 \be
 \rho(\om) = \frac{N_c^2}{16\pi}
 \frac{\om^2\sinh(\om/2T)}{\cosh(\om/2T) - \cos(\om/2T)}.
 \ee 
 These spectral functions are shown in Figs.\ \ref{figeta} and \ref{figsig}
(right). Rather than dividing by powers of $N_c$, I have simply put
$N_c=3$. It is clear that at strong coupling the transport peak is no
longer separated from the high-energy contribution and the spectral
functions go smoothly to $\om=0$. The intercept is of course proportional
to the shear viscosity and the conductivity respectively.

Comparing the spectral functions on the left with those on the right makes 
it interesting to speculate what happens in QCD in the strong coupling 
regime above the deconfinement transition.

\section{Transport from lattice QCD}

We have seen that weak-coupling methods are probably not applicable in the 
interesting temperature regime of QCD probed by current heavy ion collisions. 
Furthermore, these calculations are so involved that in most cases only 
the leading-order result is currently known; it is an open question how to 
determine subleading corrections. On the other hand, the strong-coupling 
results discussed above are obtained in theories that are not QCD. So the 
important question is what can be said about transport in QCD when 
$1\lesssim T/T_c\lesssim 3$, using nonperturbative lattice simulations.

The euclidean correlator calculated in numerical simulations is related 
to the corresponding spectral function via a dispersion relation,
 \be
 \label{eqGom}
 G_E(i\om_n) = \int_{-\infty}^{\infty} \frac{d\om}{2\pi}\,
 \frac{\rho(\om)}{\om- i\om_n},
 \ee
 where $\om_n=2\pi nT$ ($n\in \mathbb{Z}$) are the bosonic Matsubara 
frequencies. In euclidean time, with $0\leq \tau < 1/T$, this relation becomes
 \be
 \label{eqGt}
 G_E(\tau) = \int_{0}^{\infty} \frac{d\om}{2\pi}\,K(\om,\tau)\rho(\om),
 \ee
 with the kernel
 \be
 \label{eqkernel}
 K(\om,\tau) = \frac{\cosh[\om(\tau-1/2T)]}{\sinh(\om/2T)}
 = \left[1+n_B(\om)\right]e^{-\om\tau} + n_B(\om) e^{\om\tau},
 \ee
 where $n_B(\om)=1/(e^{\om/T}-1)$ is the Bose distribution.
  The first expression for the kernel shows the characteristic euclidean 
time dependence, while the second expression emphasizes that the 
correlator and its spectral function are essentially related via a Laplace 
transform, made periodic to satisfy the Kubo-Martin-Schwinger (periodicity) 
condition.

If the euclidean correlator is known analytically, the spectral function 
can simply be obtained by analytic continuation,
\be
 \rho(\om) = 2\mbox{Im}\, G_E(i\om_n\to \om+i\epsilon).
\ee
 However, if $G_E(\tau)$ is determined numerically on a lattice with
$N_\tau = 1/aT$ points in the euclidean time direction ($a$ is the
temporal lattice spacing), Eq.\ (\ref{eqGom}) has to be inverted by some
other means. This is an ill-posed inversion problem, since $G_E(\tau)$ is
known at, say, ${\cal O}(10)$ data points, whereas $\rho(\om)$ is needed
at ${\cal O}(10^3)$ values (after imposing a high-energy cutoff $\om_{\rm
max}$ and discretizing the resulting finite interval $0<\om<\om_{\rm
max}$).

One possible solution is to provide an Ansatz for the spectral function, 
with a small number of free parameters. In this case it is important to be 
able to judge the applicability of the Ansatz. An orthogonal approach is 
to avoid giving functional forms but only supply a {\em minimal} amount of 
prior information, such as positivity ($\om\rho(\om)\geq 0$) and 
asymptotic behaviour ($\rho(\om)\sim \om^n$ for large $\om$). Methods 
based on this approach are usually collectively referred to as Bayesian 
techniques.

For a weakly coupled quark-gluon plasma, the extraction of transport 
coefficients is notoriously difficult, since euclidean correlators are 
remarkably insensitive to the structure of spectral functions at energies 
$\om\ll T$ \cite{Aarts:2002cc,Petreczky:2005nh}. However, at stronger 
coupling the transport peak is much broader and the small $\om$ limit is 
no longer singular. As discussed above, smooth spectral functions are 
also found in ${\cal N} =4$ SYM at strong coupling. This opens up the 
possibility that transport coefficients are accessible in lattice QCD 
above the deconfinement transition.

So far the number of papers in the literature that have attempted to 
extract transport coefficients from the lattice in a head-on approach is 
very small. A first attempt to measure transport coefficients can be found 
in the pioneering paper by Karsch and Wyld, using an Ansatz 
\cite{Karsch:1986cq}. This method was followed by Nakamura and Sakai for 
the shear and bulk viscosity \cite{Nakamura:2004sy}. For a critical 
discussion of the Ansatz, see Ref.\ \cite{Aarts:2002cc}. S.\ Gupta used 
Bayesian methods to isolate the transport contribution at small energies 
in the case of the electrical conductivity \cite{Gupta:2003zh}. In the 
past six months, two significant steps have been made. A standard approach 
to perform the analytical continuation using Bayesian techniques is known 
as the Maximum Entropy Method (MEM), to be discussed below. It was known 
from previous work that MEM performs poorly at small energies. Aarts, 
Allton, Foley, Hands and Kim have identified and resolved a numerical 
instability in MEM in the limit that $\om\to 0$ and applied the new 
formulation to obtain the electrical conductivity \cite{Aarts:2007wj}. 
Precisely determined correlators are essential to have control over the analytic 
continuation; while for the electrical conductivity this is not a problem, 
for the shear and bulk viscosity standard measurement techniques are 
insufficient. Meyer has applied a two-level algorithm to better determine 
energy-momentum correlators and found a result for the shear viscosity 
\cite{Meyer:2007ic}.
 All calculations to date have been performed in quenched QCD, so one may 
think of the electrical conductivity in pure gauge theory as representing 
the transport properties of a single electrically charged quark diffusing 
through a gluon plasma.

 In the following I describe in some detail the Maximum Entropy Method and 
indicate why the standard algorithm is unstable in the small $\om$ region 
\cite{Aarts:2007wj}. 
In MEM one reconstructs the {\em most probable} spectral function by 
extremizing the probability distribution $P[\rho|GH]$, i.e.\ the 
probability to find $\rho$, given the correlator $G$ and prior information 
$H$. Using an identity for conditional probabilities, $P[\rho G|H] = 
P[\rho|GH] P[G|H] = P[G|\rho H] P[\rho|H]$, $P[\rho|GH]$ is written as the 
product of a standard likelihood function, $P[G|\rho H] \sim e^{-L}$ 
($\chi^2$ fit), and a prior probability, $P[\rho|H] \sim e^{\alpha S}$ 
which is independent of the data. $P[G|H]$ is a normalization factor. The 
prior information is encoded in the entropy term,
 \be
 S = \int \frac{d\om}{2\pi} \left[ \rho(\om) - m(\om) - 
 \rho(\om)\log\frac{\rho(\om)}{m(\om)}\right],
 \ee
 via the default model $m(\om)$. Therefore, the combined function to 
extremize is $P[\rho|GH] \sim e^{-L+\alpha S}$, where $\alpha$ 
determines the relative weight of the data versus the prior information 
\cite{Asakawa:2000tr}.

\begin{figure}
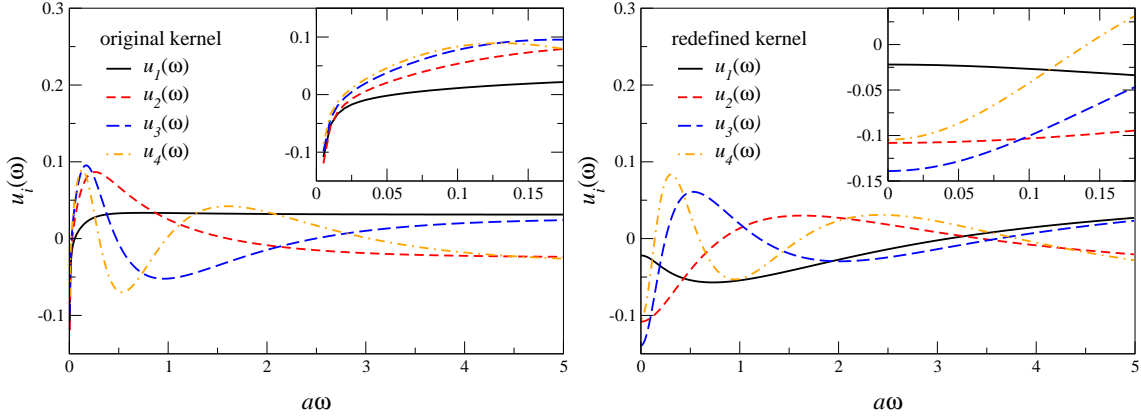

\centerline{\psfig{figure=basis_svd_ori_24_combined.eps,width=7.5cm}
\psfig{figure=basis_svd_mod_24_combined.eps,width=7.5cm}}
 \caption{First four basis functions $u_i(\om)$ in the singular value 
decomposition as a function of $a\omega$ for $a\om_{\rm max}=5$, 
$N_\om=1000$, $N_\tau=24$, using the standard (left) and the redefined 
kernel (right). The inset shows a blow-up of the small energy region.
 }
\label{figsvd}
\end{figure}

The most important aspect for my purpose here is the {\em reduction step}. 
Recall that after discretization $\rho(\om)$ is wanted at $N_\om = {\cal 
O}(10^3)$ values whereas the correlator is only known at ${\cal O}(10)$ 
points. I denote with $N$ the number of time slices included in the 
analysis; due to reflection symmetry, $N\leq N_\tau/2$. To make this 
well-defined, the number of coefficients parametrizing the spectral 
function cannot exceed $N$, or in other words, $\rho(\om)$ has to be 
restricted to an (at most) $N$ dimensional subspace. In Bryan's algorithm
\cite{Bryan}
this is achieved via a singular value decomposition (SVD) of the kernel 
$K(\om_n,\tau_i)$.  Viewed as an $N_\om\times N$ matrix, the kernel is 
written as $K = U W V^T$, where $U$ is an $N_\om\times N$ matrix, with 
$U^TU=\id_{N\times N}$, $W$ is a diagonal $N\times N$ matrix, and $V$ is 
an orthogonal $N\times N$ matrix. The $N$ dimensional subspace is spanned 
by the column vectors of $U$: $u_i(\om_n) = U_{ni}$. These basis vectors 
are orthogonal but not complete. An analysis of the extremum conditions 
shows that it is natural to write the spectral function in terms of these 
basis vectors as $\rho(\om) = m(\om) \exp \sum_{i=1}^N c_i u_i(\om)$. This 
ensures positivity and provides the actual reduction step. Extremizing the 
probability distribution leads to nonlinear equations for the $N$ 
coefficients $c_i$.

 The first four basis functions in the SVD are shown in Fig.\ \ref{figsvd}
(left), for a typical choice of $N=N_\tau/2$, $N_\om$, and $\om_{\rm
max}$. A blow-up of the small energy region reveals that the basis
functions appear to diverge when $\om\to 0$, although they are normalized
(the smallest energy included here is $a\Delta\om = a\om_{\rm max}/N_\om =
0.005$). This apparent divergence is due to the singular behaviour of the
kernel (\ref{eqkernel}): in the limit that $\om\to 0$ one finds that
$K(\om,\tau) = 2T/\om + {\cal O}\left(\om/T\right)$. Note that the leading
singular term is $\tau$ independent; all $\tau$ independence resides in
the subleading terms. The behaviour of the basis functions is therefore
indicative of a real problem, which cannot be solved by e.g.\ decreasing
$\Delta \om$. In actual applications of MEM, we (and others) found
irregular behaviour at small $\omega$, in particular at the smallest
nonzero value of $\om$.  Needless to say that this prevents access to the
transport properties encoded in euclidean correlators. It is worth
pointing out that this problem only appears at finite temperature, since
at zero temperature the kernel reduces to $K(\om,\tau)=e^{-\om\tau}$ and
the limit $\om\to 0$ is smooth. It is a manifestation of the fact that the
limits $\om\to 0$ and $T\to 0$ do not commute, which is well-known in
thermal field theory.

Once the problem is identified, it is straightforward to solve it. The 
$1/\om$ divergence can be avoided by defining
 \be
 \overline K(\om,\tau) = \frac{\om}{2T} K(\om,\tau),
 \;\;\;\;\;\;\;\;\;\;\;\;\;\;\;\;\;\;\;\;\;\;\;\;
 \overline\rho(\om) = \frac{2T}{\om} \rho(\om).
 \ee
 Since $K(\om,\tau)\rho(\om) = \overline K(\om,\tau)\overline \rho(\om)$ 
the standard relation with the euclidean correlator holds. However, the 
modified kernel is finite when $\om\to 0$: $\overline K(0,\tau)=1$. A SVD 
of $\overline K$ yields new basis functions $\overline u_i(\om)$. The 
first four are shown in Fig.\ \ref{figsvd} (right). Clearly they take a 
finite value when $\om\to 0$, and the $\om=0$ point can be included in the 
analysis. The redefined spectral function is expanded as 
$\overline{\rho}(\om) = \overline{m}(\om) \exp \sum_{i=1}^N \overline{c}_i 
\overline{u}_i(\om)$ and the same MEM routine can be used to find the 
coefficients $\overline{c}_i$. MEM now reconstructs $\overline\rho \sim 
\rho/\om$ rather than $\rho$. This reshuffling of powers of $\om$ is 
nontrivial, since $\overline\rho$ and $\rho$ are not expanded in a 
complete set: the reduction step restricts $\overline\rho$ to a different 
subspace with manifestly different properties, in particular in the small 
$\om$ limit.

 A second (minor) modification is needed to access $\rho(\om)/\om$ at zero
$\om$, relevant for transport coefficients. For spectral functions of
fermion bilinears, such as $j^\mu=\bar\psi\gamma^\mu\psi$, the traditional
default model is $\overline{m}(\om)\sim m(\om)/\om \sim \om$, determined
by the high-energy behaviour $\rho(\om) \sim\om^2$. Unfortunately, this
introduces a bias and puts the intercept equal to zero from the start. To
avoid this, one may use $\overline{m}(\om)\sim (b+\om)$, where $b>0$ is a
parameter that can be used to assess default model dependence at small
$\om$.

\begin{figure}
 \centerline{\epsfig{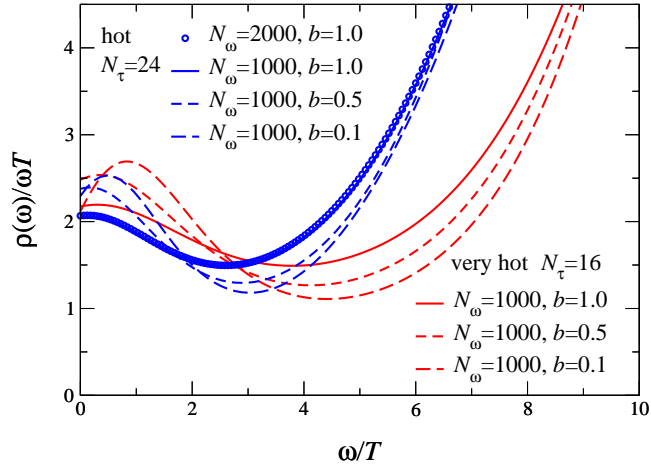}}
 \caption{
 Current-current spectral function $\rho(\om)/\om T$, where 
$j^i=\bar\psi\gamma^i\psi$, as a function of $\om/T$ in quenched QCD for 
$T/T_c\sim 1.5$ (hot, $N_\tau=24$) and 2.25 (very hot, $N_\tau=16$). The 
intercept at $\om=0$ is proportional to the electrical conductivity. 
Results are shown for $N_\om=1000$, $2000$ and 
$b=1.0$, $0.5$, $0.1$ at fixed $a\om_{\rm max}=5$.
 }
 \label{fig:rho_zoom}
\end{figure}

We have applied the modified algorithm to the problem of the electrical 
conductivity (or charge diffusion) in quenched QCD with light staggered 
fermions and performed simulations on a fine lattice at $\beta=7.192$ of 
size $64^3\times N_\tau$ above the deconfinement transition. The quarks 
are so light that chiral symmetry restoration is clearly visible when 
comparing pseudoscalar and scalar correlators \cite{Aarts:2006wt}. In 
Fig.\ \ref{fig:rho_zoom}, the spectral function normalized by $\om T$ is 
shown for not too large $\om$ at two temperatures above the deconfinement 
transition: $T/T_c\sim 1.5$ ($N_\tau=24$) and $T/T_c\sim 2.25$ 
($N_\tau=16$). In both cases results for three values of the default model 
parameter $b$ are shown. The slight spread of the curves gives an 
indication of the uncertainty in the MEM reconstruction. There is no 
dependence on the discretization along the $\om$ axis, as can be seen from 
the results at $N_\tau=24$ using $N_\om=1000$ and $2000$ at fixed 
$\om_{\rm max}$. From the intercept the conductivity is found to be 
$\sigma/T = 0.4 \pm 0.1$ with no significant temperature dependence. This 
result is normalized to a single flavour and should be multiplied with the 
sum of the electric charge squared for light flavours. The error is 
systematic and due to the MEM uncertainty. The statistical error is 
expected to be smaller. This result is indicative of strong interactions: 
at weak coupling the conductivity behaves as $\sigma/T\sim 1/(g^4\ln 1/g) 
\to \infty$, whereas at strong coupling the scale is set solely by the 
temperature.

A similar conclusion has been drawn by Meyer in the case of the shear 
viscosity. In Ref.~\cite{Meyer:2007ic} an upper bound $\eta/s\lesssim 1$ 
is obtained in SU(3) gauge theory on lattices at $T/T_c = 1.24$ 
($\beta=6.2$, $20^3\times 8$) and 1.65 ($\beta=6.408$, $28^3\times 8$).

\section{Quarkonium at high temperature}

Another signal of strong interactions in the quark-gluon plasma is the 
survival of charmonium and other heavy quark mesons above $T_c$. This has 
been studied on the lattice a few years ago using the Maximal Entropy 
Method \cite{Asakawa:2003re,Datta:2003ww,Umeda:2002vr}. A more recent 
extensive study of charmonium and bottonium spectral functions can be 
found in Ref.\ \cite{Jakovac:2006sf}. This topic has been discussed last 
year in Hatsuda's plenary talk \cite{Hatsuda}, so I briefly mention 
developments that took place recently and were discussed at this 
Conference.

Up to last year, all studies were performed in quenched QCD. The TrinLat
collaboration has carried out dynamical simulations with two flavours on
highly anisotropic lattices, which can be used to study charmonium at zero
and nonzero temperature \cite{Morrin:2006tf}. A spectral function analysis
above $T_c$, using the modified version of MEM described in the previous
section, can be found in Ref.\ \cite{Aarts:2007pk}. The results suggest
that the S-waves ($J/\psi$ and $\eta_c$) survive up to temperatures close
to $2T_c$, while the P-waves ($\chi_{c0}$ and $\chi_{c1}$) melt away below
$1.2T_c$. However, there are systematic uncertainties that need to be
improved in order to make these conclusions more firm, in particular
simulations at a finer lattice spacing would be desirable. One reason this
is necessary is to better understand the appearance of artefacts at larger
energy introduced by the finite lattice spacing, which have been discussed
for a variety of lattice fermion formulations (Wilson, staggered, domain
wall, overlap) in the free field limit
\cite{Karsch:2003wy,Aarts:2005hg,Aarts:2006em}.

The presence of the transport contribution at small $\om$ can interfere 
with spectral features at larger $\om$ when not properly disentangled, as 
emphasized by Umeda \cite{Umeda:2007hy}. This can be partially avoided by 
subtracting the midpoint value at $\tau=1/2T$ from the correlator, which 
has the effect of suppressing the contribution at small $\om$. A spectral 
analysis can then be done on $G(\tau)-G(1/2T)$.

Petreczky and M\'ocsy have provided a closer look at potential models, 
traditionally used to study quarkonium at zero temperature. Doubts whether 
potential models can describe quarkonium correlators at finite temperature 
have been expressed in Ref.\ \cite{Mocsy:2005qw}. A lower melting 
temperature than usually obtained with MEM is found when a potential model 
based on lattice QCD simulations is used, both in the quenched 
approximation \cite{Mocsy:2007yj} as well as in the theory with $2+1$ flavours 
\cite{Mocsy:2007jz}. 

In order to avoid conceptual problems with the extension of
zero-temperature potential models to nonzero temperature, a real-time
static potential, firmly based in thermal field theory, was introduced in
Ref.\ \cite{Laine:2006ns}. When applied to quarkonium \cite{Laine:2007gj},
the results seem to support the standard interpretation of results
obtained with MEM.

\section{Summary}

Results from relativistic heavy ion collisions at RHIC have highlighted
the importance of understanding transport and hydrodynamical behaviour in
QCD above the deconfined transition.  Nonperturbative first-principle
calculations of spectral functions, especially at small energies
$\om\lesssim T$, are badly needed. Since this involves inherently
real-time physics, it is a difficult problem for lattice QCD, but recently
several steps forward have been made. Using multi-level algorithms,
accurately determined euclidean correlators of the energy-momentum tensor
are now available. Concerning the analytical continuation to real time, an
instability at small energies in the standard Maximum Entropy Method,
preventing access to hydrodynamical features of spectral functions, has
been found and resolved. The first results support the idea that the
quark-gluon plasma is strongly interacting in the temperature range $1
\lesssim T/T_c \lesssim 2.5$. Extension of the work described here will
hopefully yield a better understanding of the hydrodynamical regime of
thermal QCD from first principles.

\section*{Acknowledgments}

It is a great pleasure to thank my collaborators Chris Allton, Justin
Foley, Simon Hands and Seyong Kim, Bugra Oktay, Mike Peardon
and Jonivar Skullerud, and Jose Mart{\'\i}nez Resco.
 I also thank Carlos Nunez for discussion.
 This work is supported by a PPARC Advanced Fellowship.

\end{document}